\documentclass[usenatbib,useAMS]{mn2e}
\title[Keck spectroscopy of CLASS gravitational lenses]{Keck spectroscopy of CLASS gravitational lenses}
\author[J. P. McKean et al.]{J. P. McKean$^{1,2}$\thanks{E-mail: mckean@physics.ucdavis.edu}, L. V. E. Koopmans$^{3,4}$, I. W. A. Browne$^{1}$, C. D. Fassnacht$^{2}$, 
\newauthor R. D. Blandford$^{5}$, L. M. Lubin$^{2}$ and A. C. S. Readhead$^{3}$\\
$^{1}$University of Manchester, Jodrell Bank Observatory, Macclesfield, SK11 9DL, U.K.\\
$^{2}$Department of Physics, University of California, Davis, CA 95616, U.S.A\\
$^{3}$California Institute of Technology, Pasadena, CA 91125, U.S.A.\\
$^{4}$STScI, 3700 San Martin Dr., Baltimore, MD 21218, U.S.A.\\
$^{5}$KIPAC, Stanford University, 2575 Sand Hill Road, Menlo Park, CA 94025, U.S.A.}
\begin{document}

\date{Accepted 2004 January 16. Received 2004 January 14; in original form 2003 December 2}

\pagerange{\pageref{firstpage}--\pageref{lastpage}} \pubyear{2002}

\maketitle

\label{firstpage}

\begin{abstract}
We present the optical spectra of four newly discovered gravitational lenses from the Cosmic Lens All-Sky Survey (CLASS). These observations were carried out using the Low Resolution Imaging Spectrograph on the W. M. Keck-I Telescope as part of a program to study galaxy-scale gravitational lenses. From our spectra we found the redshift of the background source in CLASS B0128+437 ($z_s=3.1240\pm0.0042$) and the lensing galaxy redshifts in CLASS B0445+123 ($z_l =0.5583\pm0.0003$) and CLASS B0850+054 ($z_l=0.5883\pm0.0006$). Intriguingly, we also discovered that CLASS B0631+519 may have two lensing galaxies ($z_{l,1}=0.0896\pm0.0001$, $z_{l,2}=0.6196\pm0.0004$). We also found a single unidentified emission line from the lensing galaxy in CLASS B0128+437 and the lensed source in CLASS B0850+054. We find the lensing galaxies in CLASS B0445+123 and CLASS B0631+519 ($l,2$) to be early-type galaxies with Einstein Radii of $2.8-3.0~h^{-1}$~kpc. The deflector in CLASS B0850+054 is a late-type galaxy with an Einstein Radius of $1.6~h^{-1}$~kpc.
\end{abstract}


\section{Introduction}
Gravitational lensing has the ability to probe the internal mass distributions of galaxies at cosmological distances. The ability to form complete samples, based solely on mass, has allowed studies of the formation and evolution of early-type galaxies at intermediate redshifts (e.g. \citealt*{keeton98,treu02,koopmans03}; \citealt{rusin03}). Anomalous flux-density ratios observed in merging gravitational lens images have recently been used to argue for cold dark matter substructure within the haloes of lensing galaxies \citep{metcalf01,keeton01,chiba02,metcalf02,dalal02,bradac02,keeton03}. The magnification of the lensed source has also allowed studies of star-formation at high redshift (e.g. \citealt{ebbels96,ellis01}) and the dust content of quasar host galaxies to be estimated \citep{barvainis02}. Furthermore, gravitational lenses have proved to be powerful tools for determining the cosmological parameters. Lenses with well-constrained mass models and accurate time-delay measurements have been used to calculate the Hubble parameter (e.g. \citealt{schechter97,lovell98,biggs99,fassnacht99,koopmans99,fassnacht02a,treu02a}). The gravitational lensing statistics from large systematic surveys have provided complementary and independent constraints on the cosmological constant and density parameters (\citealt{kochanek96a}; \citealt*{falco98}; \citealt{helbig99,chae02,chae03}) to those obtained from SN1a \citep{riess98,perlmutter99}, large-scale structure \citep{percival01,tegmark02} and cosmic microwave background \citep{slosar03,sievers03,spergel03} observations. In addition, the nature of the curvature parameter can also be determined from the image-separations and source redshifts via the $\Delta\theta$-$z_{s}$ relation \citep*{turner84,gott89,park97,williams97,helbig98}. However, each of these applications of gravitational lensing is critically dependent on the redshifts of the lensing galaxy and lensed source being known. 

With the objectives outlined above in mind, a spectroscopic survey of galaxy-scale gravitational lenses using the W. M. Keck and Palomar Observatories has been underway over the last few years \citep{kundic97,fassnacht98,lubin00}. In this paper we present our latest results. The observing sample consisted of four gravitational lenses discovered during the course of the Cosmic Lens All-Sky Survey (CLASS; \citealt{myers02,browne02}). In Section \ref{spec-targets} a short review of each of these gravitational lenses is given. The observations with the W. M. Keck-I Telescope are presented in Section \ref{spec-obs} and the optical spectra are presented in Section \ref{spec-results}. The resulting implications for each gravitational lens are discussed in Section \ref{spec-disc}. Finally, in Section \ref{spec-conc} a summary of the observations and analysis presented in this paper is given. Throughout we adopted an $\Omega_{M}=0.3$, $\Omega_{\Lambda}=0.7$ flat-universe, with a Hubble parameter $H_{0}=100~h$~km~s$^{-1}$~Mpc$^{-1}$.

\section{Targets}
\label{spec-targets}

The targets of our observations were CLASS B0128+437, CLASS B0445+123, CLASS B0631+519 and CLASS B0850+054. A short review of each gravitational lens is given below.

\subsection{CLASS B0128+437}

The quadruple-imaged gravitational lens system CLASS B0128+437 is described in \citet{phillips00b} and \citet{biggs03b}. MERLIN (Multi Element Radio Linked Interferometer Network) 5~GHz imaging detected four compact components, with a maximum image separation of $0.54\arcsec$, arranged in a classic quad-lens formation. The 325~MHz to 8.46~GHz radio-spectrum was found to be strongly peaked around $\sim~1$~GHz, which suggested the lensed source may be a Compact Symmetric Object (CSO; e.g. \citealt{readhead96}). Follow-up VLBA (Very Long Baseline Array) imaging at 5~GHz found sub-structure consistent with the lensing of a CSO hypothesis. However, it should be noted that the recent VLBI observations of \citeauthor{biggs03b} have placed the CSO interpretation in some doubt since a flat-spectrum compact core appears to be present in 3 of the 4 images, suggesting a core-jet structure. Optical HST (Hubble Space Telescope) and infra-red UKIRT (United Kingdom Infra-Red Telescope) photometry found CLASS B0128+437 to be highly reddened ($I-K\sim6$), with the dominant infra-red emission being from the lensed images. Therefore, the background source in CLASS B0128+437 is thought to be highly redshifted and/or the lensing galaxy is extremely dusty. Recent modelling attempts favour lensing by a spiral galaxy, although conclusive observational evidence of this has still to be obtained \citep{norbury02}.

\subsection{CLASS B0445+123}

CLASS B0445+123 (\citealt{argo03}) is a double-imaged  gravitational lens system. 8.46~GHz VLA (Very Large Array) and 5~GHz MERLIN and VLBA observations found two compact components separated by $1.32\arcsec$ with identical radio-spectra. More sensitive 8.46~GHz VLA imaging detected extended, arc-like structure in component A. Imaging with the WHT (William Herschel Telescope) detected faint extended optical emission with an $R$-band magnitude of $21.8\pm0.4$-mag. However, the resolution was not sufficient to separate the lensed images from the lensing galaxy.

\subsection{CLASS B0631+519}

CLASS B0631+519 (York et al., in preparation) is a new double-imaged gravitational lens system discovered by CLASS. This lens was found to have two compact components separated by $1.16 \arcsec$ when observed at 8.46~GHz with the VLA. Observations with MERLIN and the VLBA at 5~GHz detected structure in the radio components which was consistent with gravitational lensing. Optical imaging with the WHT detected emission ($R$-band $21.4\pm0.1$-mag) mainly from the lensing galaxy. The wealth of observational constraints available in CLASS B0631+519 make this gravitational lens an exciting prospect for detailed modelling of the lensing potential in the future.

\subsection{CLASS B0850+054}

CLASS B0850+054 \citep{biggs03} is a double-imaged gravitational lens system. The two components, separated by $0.68\arcsec$, are compact when observed with the VLA and MERLIN at 8.46~GHz and 5~GHz respectively. However, high resolution imaging with the VLBA at 5~GHz shows sub-structure in component A consistent with gravitational lensing. UKIRT $K$-band imaging detected infra-red emission (18.4-mag) believed to be mainly from the lensing galaxy. Unfortunately, the resolution was insufficient to determine the position of the lensing galaxy relative to the lensed images and mass-modelling has not been possible. A flux-density monitoring program at 8.46~GHz with the VLA has not detected any variability in CLASS B0850+054. Like CLASS B0128+437, the radio-spectrum of CLASS B0850+054 turns-over at $\sim 1$~GHz which, coupled with the lack of variability, suggests the background source may be a CSO.

\begin{figure*}
\begin{center}
\setlength{\unitlength}{1cm}
\begin{picture}(15,5.5)
\put(-1.1,-0.05){\includegraphics{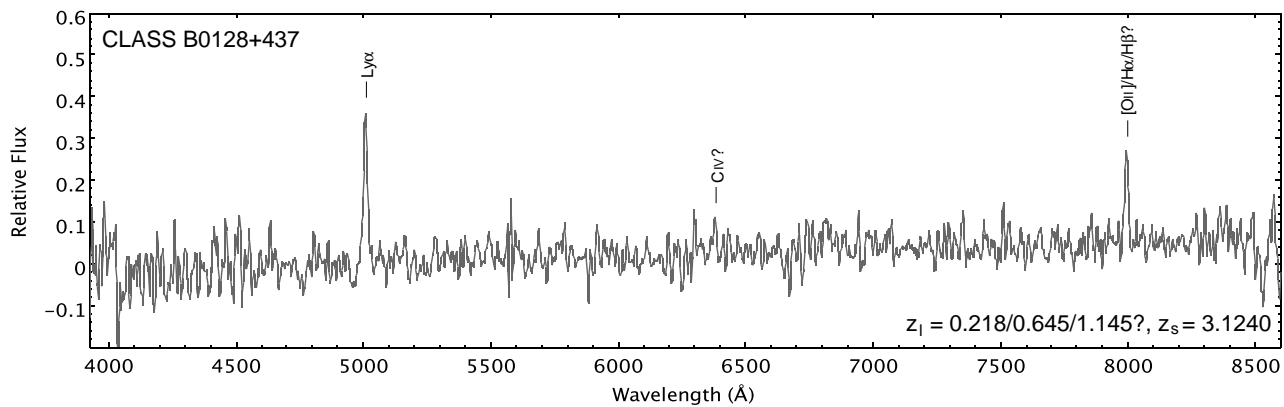}}
\end{picture}
\caption{The optical spectrum of CLASS B0128+437 taken with the W. M. Keck-I Telescope on 17th November 2001 using LRIS. Three emission lines have been detected; two from the source, Ly$\alpha$ at 5015~{\AA} and possibly C~{\sc iv} at 6383~{\AA}, and one from the lens at 7995~{\AA}, which is believed to be [O~{\sc ii}], H$\beta$ or H$\alpha$. These correspond to a lens at $z_{l}=1.145/0.645/0.218$ with the background source at $z_{s} = 3.1240\pm0.0042$. The relative flux has units of $\times 10^{-17}$~erg~cm$^{-2}$~s$^{-1}$~{\AA}$^{-1}$. The spectrum has been smoothed using a boxcar of 10~{\AA}.}
\label{lens-0128}
\end{center}
\end{figure*}

\section{Observations \& Data Reduction}
\label{spec-obs}

The spectroscopic observations were carried out using the W. M. Keck-I 10-m Telescope on 2001 November 16 and 17. On both nights the conditions were photometric and the FWHM seeing was typically $\sim 0.5 \arcsec - 0.7\arcsec$. The optical spectroscopy was taken through the Low Resolution Imaging Spectrograph (LRIS; \citealt{oke95}) with a $1 \arcsec$-wide longslit. LRIS has a red (LRIS-R) and a blue\footnote{Note that these observations were carried out with the `old LRIS blue' camera.} (LRIS-B) camera; each with a $2048 \times 2048$ pixel array. Together they can provide spectral coverage between $3000-10000$~{\AA}. The plate-scale of both arrays is $0.215 \arcsec$~pixel$^{-1}$ and the field of view is $ \sim 6 \arcmin \times 8 \arcmin$. The light beam was split between LRIS-B and -R using a 6800~{\AA} dichroic. A $300$~grooves~mm$^{-1}$ grism blazed at 5000~{\AA} produced a dispersion of $2.32$~{\AA}~pixel$^{-1}$ and a wavelength coverage of $\sim 4000-6800$~{\AA} through LRIS-B. A $400$~grooves~mm$^{-1}$ grating blazed at $8500$~{\AA} and centred at $8200$~{\AA} was used through LRIS-R. This had a dispersion of $\sim 1.86$~{\AA}~pixel$^{-1}$ and a wavelength coverage of $\sim 6800-9100$~{\AA}. Throughout the observing run LRIS-B and -R were set to single and dual amp readout modes, respectively. Observations of the spectrophotometric standard star BD$+75\degr325$ \citep{oke90} were taken to flatten the spectra and identify sky absorption features during the reduction stage. At the end of each night, spectroscopic dome flat-field exposures, bias frames and arcs were taken. 

During the first night the LRIS autoguider and rotator were not operational. As a consequence imaging and accurate guiding were not possible. Spectroscopy of CLASS B0631+519 was therefore attempted as it was optically bright enough for accurate manual guiding. The rotator was functioning on the second night which allowed accurate offsetting to the fainter gravitational lenses. However, slit position angles were set such that each gravitational lens and a suitably bright nearby object were in the slit. This allowed confirmation that each gravitational lens had been detected. A summary of the observing run is given in Table \ref{log}. The exposure times were typically $n \times 1800$~seconds. However, for CLASS B0850+054 the exposure times for LRIS-B and -R differed due to a software problem with the LRIS-B control system during the second observation.

\begin{table}
\begin{center}
\begin{tabular}{ccccr} \hline
Gravitational   & Date		& \multicolumn{2}{c}{Exposure Time ($s$)}	& \multicolumn{1}{c}{PA}  \\ 
Lens		&		& LRIS-B	& LRIS-R		& \multicolumn{1}{c}{($\degr$)} \\ \hline
CLASS B0128+437	& 17Nov01	& 1800		& 1800			& 42.2     \\
CLASS B0445+123	& 17Nov01	& 3600		& 3600			& 333.6    \\
CLASS B0631+519 & 16Nov01	& 5100		& 5100			& n/a        \\
CLASS B0850+054 & 17Nov01	& 1800		& 4200			& 75.0     \\ \hline
\end{tabular}
\caption{Observing logs for spectroscopic observations of CLASS gravitational lenses CLASS B0128+437, CLASS B0445+123, CLASS B0631+519 and CLASS B0850+054. Slit position angles (PA) are given east of north. For the  CLASS B0631+519 observation there was no defined slit PA because the instrument rotator was not operational and therefore the slit PA was not fixed.}
\label{log}
\end{center}
\end{table}

The data were reduced in the standard manner using {\sc iraf}\footnote{{\sc iraf} ({\sc i}mage {\sc r}eduction and {\sc a}nalysis {\sc f}acility) is distributed by the National Optical Astronomy Observatories, which are operated by AURA, Inc., under cooperative agreement with the National Science Foundation.}. The bias level was estimated using the overscan regions on each image and subtracted. The individual spectral flat-field frames were combined by median filtering and  normalised by fitting a cubic-spline fourth order polynomial function before being applied to the data. Cosmic-rays were removed from each exposure using {\sc lacosmic} \citep{dokkum01} and multiple exposures of each gravitational lens were combined. The wavelength calibration was applied using exposures of Hg, Ar and Ne arc lamps. However, during this process a systematic shift in wavelength of $\la 5$~{\AA} was found. This shift is almost certainly due to flexure in the instrument, which is a function of elevation and rotation. As the arc lamp exposures were taken at the end of the night and not directly after each exposure, there was a small systematic shift in the wavelength scale. This was corrected for using the positions of the sky-lines. The one-dimensional spectra were optimally extracted in the variance weighted mode. The apertures were set to $\sim 15$~pixels which corresponded to $\sim 3$ times the seeing and therefore should have contained most of the signal from each lens. Apertures of similar size were placed above and below the target aperture to estimate the sky background contribution. A simple second order Legendre function was used to trace the continuum of BD$+75\degr325$. This trace was then used as a reference during the extraction of the weaker gravitational lens spectra. Using the spectrophotometric standard star spectrum, the extracted one-dimensional spectra of each gravitational lens were flux calibrated and flattened. The LRIS-B and -R data were then combined to produce the final calibrated spectra which ranged from $4000 - 9100$~{\AA}.

\section{Results}
\label{spec-results}

The calibrated one-dimensional optical spectrum of each gravitational lens was analysed using the {\sc splat}\footnote{Part of the {\sc starlink} Project - http://www.starlink.rl.ac.uk} ({\sc sp}ectra{\sc l} {\sc a}nalysis {\sc t}ool) package. The centroid of each emission/absorption feature was measured by fitting Gaussian line profiles to the un-smoothed continuum-subtracted spectra. The redshift of each lensing galaxy/lensed source was determined from the unweighted mean redshift, $\langle z \rangle$, of the observed emission and absorption features. The corresponding redshift uncertainties were calculated from the rms scatter in the emission/absorption feature redshifts. Note that these uncertainties do not account for any systematic uncertainties introduced by the instrument flexure or the Gaussian fitting procedure. A short description of the spectrum obtained for each gravitational lens is given below.

\subsection{CLASS B0128+437}

\begin{figure}
\begin{center}
\setlength{\unitlength}{1cm}
\begin{picture}(6,5)
\put(-1.1,-0.05){\includegraphics{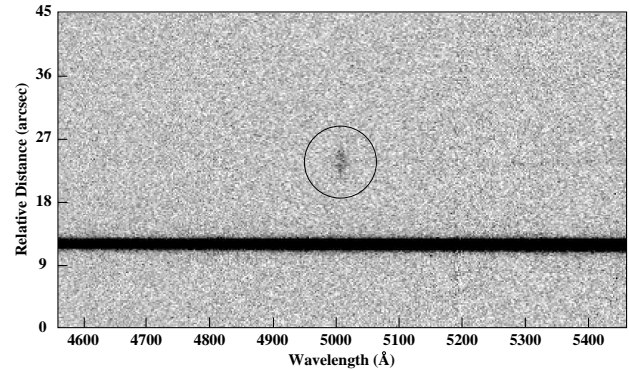}}
\end{picture}
\caption{A portion of the background subtracted 2-d LRIS-B spectrum of CLASS B0128+437 between 4565 and 5465~{\AA}, centred on 5015~{\AA}. The circled emission line, which is at 5015~{\AA} is almost certainly Ly$\alpha$. The bright trace is due to the offset star J0131+4358. The range of the spatial axis is $45 \arcsec$.}
\label{lens-0128-bac}
\end{center}
\end{figure}

The optical spectrum of CLASS B0128+437, presented in Fig. \ref{lens-0128}, shows two strong emission lines at 5015 and 7995~{\AA} with very little surrounding continuum. Fig. \ref{lens-0128-bac} shows a portion of the background subtracted two-dimensional spectrum of CLASS B0128+437 between 4565 and 5465~{\AA}, centred on 5015~{\AA}. There is clearly a strong and extended (by $\sim 3.2\arcsec$ in the spatial axis) emission line in the CLASS B0128+437 spectrum at 5015~{\AA}. Toward the red end of this line there is evidence of faint continuum which is not present toward the blue end. Intriguingly, this emission line appears to have a small gap where the continuum should be. This, coupled with the line profile, leads to the conclusion that this line is almost certainly Ly$\alpha$ from the background source at a redshift of $z_{s} = 3.1240\pm0.0042$. Consistent with this redshift is the tentative detection ($\la 2 \sigma$) of C~{\sc iv} emission at 6383~{\AA}. The other strong emission line at 7995~{\AA} is not associated with the $z = 3.1240$ background source, but is presumably from the lensing galaxy. This emission line is strong and isolated and is likely to be [O~{\sc ii}], H$\alpha$ or possibly H$\beta$. However, as there was only very faint continuum detected, it has not been possible to identify any absorption features in the lensing galaxy spectrum which could confirm one of these hypotheses. Attempts, encouraged by the results presented here, have been made by Biggs (private communication) to detect H~21-cm absorption using the Westerbork Synthesis Radio Telescope. Thus far, these observations have been unsuccessful, but are ongoing. The results from Gaussian profile fitting to each line are presented in Table \ref{spec-red-tab}. 

\subsection{CLASS B0445+123}

\begin{figure*}
\begin{center}
\setlength{\unitlength}{1cm}
\begin{picture}(15,5.5)
\put(-1.1,-0.05){\includegraphics{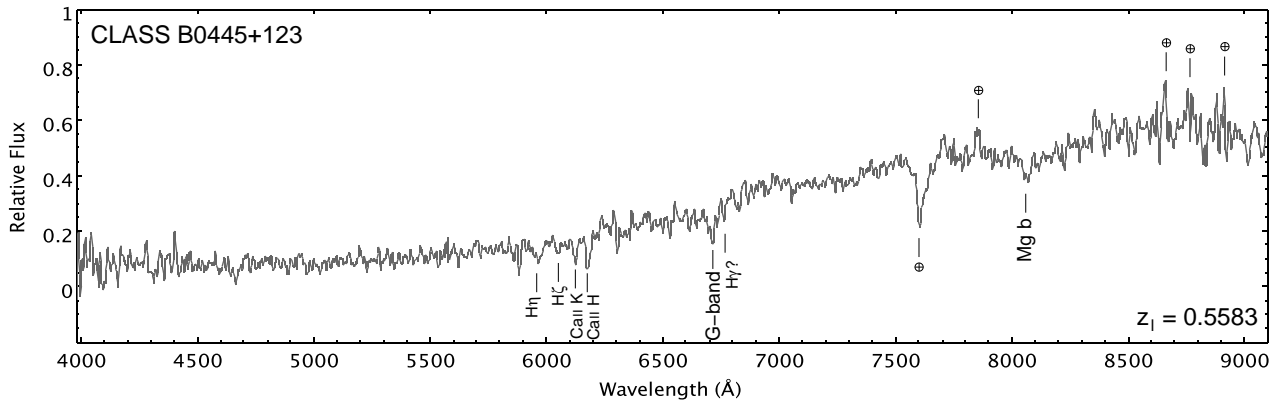}}
\end{picture}
\caption{The optical spectrum of CLASS B0445+123 taken with the W. M. Keck-I Telescope on 17th November 2001 using LRIS. Seven absorption lines in the lensing galaxy spectrum have been detected giving a redshift of $z_{l} = 0.5583\pm0.0003$. The broad appearance of the H$\eta$ absorption is probably due to the presence of a blend, which also includes the Mg~{\sc i} and Fe~{\sc i} absorption lines. The lensing galaxy in CLASS B0445+123 has the spectral shape of an early-type galaxy (c.f. NGC3379; \citealt{santos01}). The relative flux has units of $\times 10^{-17}$~erg~cm$^{-2}$~s$^{-1}$~{\AA}$^{-1}$. The spectrum has been smoothed using a boxcar of 10~{\AA}.}
\label{lens-0445}
\end{center}
\end{figure*}

The optical spectrum of CLASS B0445+123 is presented in Fig. \ref{lens-0445} and shows strong continuum with seven absorption features which we associate with the lensing galaxy. The results of Gaussian line profiles fitted to each feature, presented in Table \ref{spec-red-tab}, give a lensing galaxy redshift of $z_{l} = 0.5583\pm0.0003$. Note that the absence of H$\beta$ absorption at this redshift is due to the atmospheric absorption band at $\sim 7600$~{\AA}. The overall shape of the lensing galaxy spectrum is consistent with that of a passively evolving early-type galaxy (c.f. with the spectrum of NGC 3379; \citealt{santos01}) and the lack of any [O~{\sc ii}] emission indicates very little star-formation. The intensity of the 4000~{\AA} break discontinuity of early- and late-type galaxies is a good indicator of the contamination of the stellar spectrum from the presence of an active nucleus or a background source. Studies of the 4000~{\AA} break contrast in the spectra of passively evolving luminous early-type galaxies have found $D_{4000} \sim 0.5\pm0.05$ \citep{dressler87}. Those galaxies which also have emission from a nuclear component have in general discontinuity contrasts of $D_{4000} \la 0.4$ \citep{marcha96a}. It should also be noted that recent bursts of star-formation will reduce the 4000~{\AA} break contrast \citep{kimble89}. For CLASS B0445+123 the 4000~{\AA} break discontinuity was found to be $D_{4000} \sim 0.35$. The extra flux present in the CLASS B0445+123 spectrum is probably not due to an active nucleus within the lensing galaxy as there is no evidence of emission lines in the spectrum or radio emission in the \citeauthor{argo03} 5~GHz MERLIN image. The most likely source of this extra emission is from the lensed object. However, no emission lines associated with the lensed source were detected. Therefore, the source redshift remains unknown.

\subsection{CLASS B0631+519}

\begin{figure*}
\begin{center}
\setlength{\unitlength}{1cm}
\begin{picture}(15,5.5)
\put(-1.1,-0.05){\includegraphics{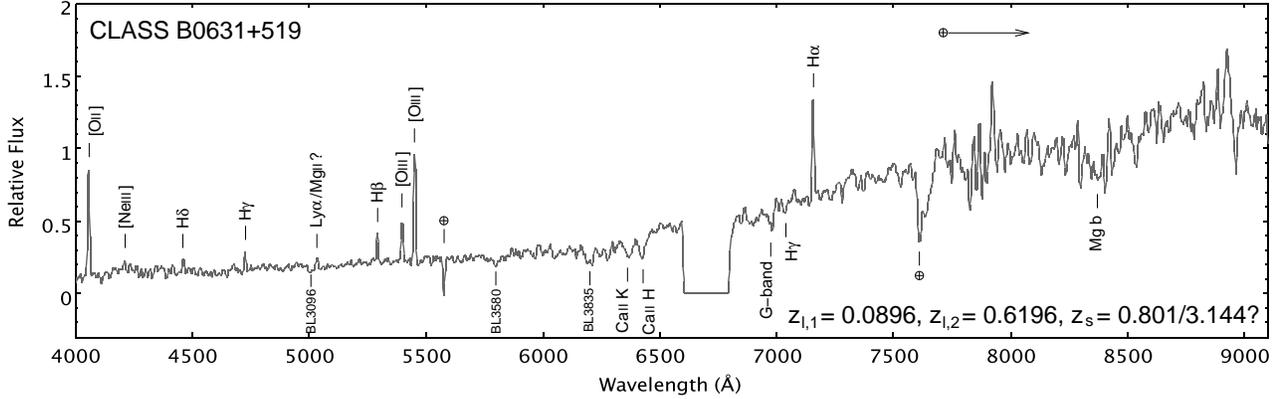}}
\end{picture}
\caption{The optical spectrum of CLASS B0631+519 taken with the W. M. Keck-I Telescope on 16th November 2001 using LRIS. Eight emission lines from a lensing galaxy at $z_{l,1}=0.0896\pm0.0001$ and seven absorption features from a second lensing galaxy at $z_{l,2}=0.6196\pm0.0004$ have been detected. The redshift of the background source is unknown although a weak emission line at 5040~{\AA} may be either Ly$\alpha$ or Mg~{\sc ii}. The data above 7700~{\AA} are very noisy and are included only to show that the second deflector is an early-type galaxy. The relative flux has units of $\times 10^{-17}$~erg~cm$^{-2}$~s$^{-1}$~{\AA}$^{-1}$. The spectrum has been smoothed using a boxcar of 10~{\AA}.}
\label{lens-0631}
\end{center}
\end{figure*}

The optical spectrum of CLASS B0631+519 is presented in Fig. \ref{lens-0631}. During the reduction of this spectrum the data above $\lambda \geq 7700$~{\AA} were found to be very noisy. This was probably due to inadequate flat-fielding which failed to remove fringing effects at longer wavelengths and poor tracking throughout the observation (between the first and third observation of CLASS B0631+519 the source had moved $1\arcsec$ on the slit). The data above $\lambda \geq 7700$~{\AA} are only included in Fig. \ref{lens-0631} for spectral classification purposes and any emission/absorption features above $\lambda \geq 7700$~{\AA} are not to be trusted. 

The CLASS B0631+519 spectrum shows strong continuum with absorption and emission lines at two very different redshifts. All of the strong emission lines come from a single source at $z_{l,1}=0.0896\pm0.0001$. The absorption lines and 4000~{\AA} break are associated with a source at $z_{l,2}=0.6196\pm0.0004$. The results of fitting Gaussian profiles to each feature are given in Table \ref{spec-red-tab}. Both objects are believed to be lensing galaxies as it is unlikely that the low redshift galaxy is capable of producing $1.16\arcsec$ image splitting of a background source at $z = 0.6196$ (see Section \ref{spec-disc} for full discussion). The absorption features and 4000~{\AA} break discontinuity contrast of $D_{4000} \sim0.4$ indicate that most of the continuum is from the second deflector, whose spectral shape is consistent with a passively evolving early-type galaxy. Although the break contrast suggests there is some additional flux in the spectrum (which may be due to the nearby lensing galaxy), no obvious emission lines were detected from the background source. Intriguingly, there is a weak peak at 5040~{\AA} which could be due to Ly$\alpha$ or Mg~{\sc ii} emission from the lensed object. Unfortunately, there is no other supporting evidence of this identification and the background source redshift remains unknown.

\subsection{CLASS B0850+054}

\begin{figure*}
\begin{center}
\setlength{\unitlength}{1cm}
\begin{picture}(15,5.5)
\put(-1.1,-0.05){\includegraphics{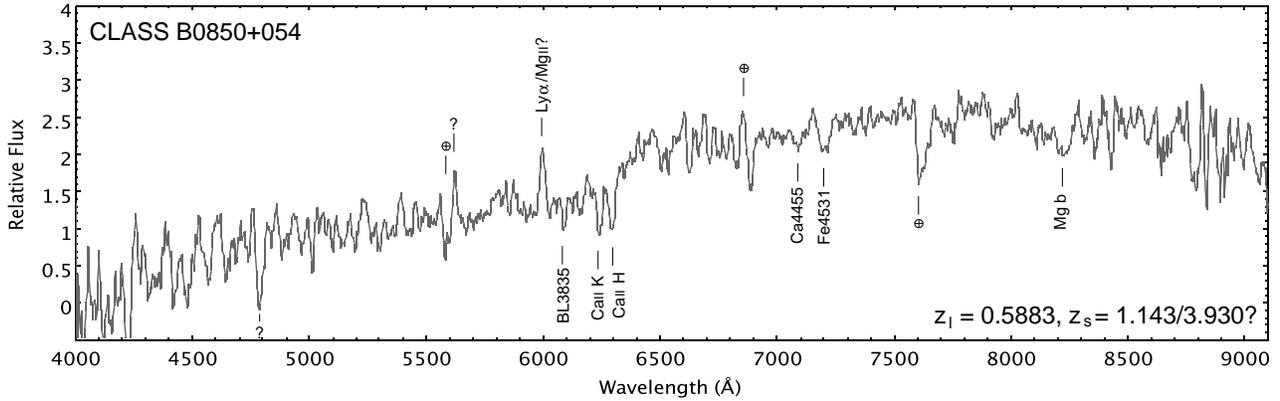}}
\end{picture}
\caption{The optical spectrum of CLASS B0850+054 taken with the W. M. Keck-I Telescope on 17th November 2001 using LRIS. Six absorption features and the 4000~{\AA} break give a lensing galaxy redshift of $z_{l}=0.5883\pm0.0006$. One strong emission line has been detected at 5995~{\AA} which has been tentatively classified as either Ly$\alpha$ or Mg~{\sc ii}. The relative flux has units of $\times 10^{-18}$~erg~cm$^{-2}$~s$^{-1}$~{\AA}$^{-1}$. The spectrum has been smoothed using a boxcar of 18~{\AA}.}
\label{lens-0850}
\end{center}
\end{figure*}

The spectrum of CLASS B0850+054 is presented in Fig. \ref{lens-0850}. Six absorption features and the 4000~{\AA} break have been detected in the lensing galaxy spectrum which was found to have a redshift of $z_{l}=0.5883\pm0.0006$. The shape of the spectrum is consistent with a late-type galaxy (Sb; c.f. with the spectrum of NGC 3627; \citealt{santos01}) and the break discontinuity is $D_{4000} \sim 0.4$ which suggests there is little contamination from any light from the background source. There is a strong emission feature in the spectrum at 5995~{\AA} which is inconsistent with the $z_{l}=0.5883$ lensing galaxy redshift. Hypothesising that this line is either Ly$\alpha$ or Mg~{\sc ii} results in a source redshift of $z_{s}=3.930$ or $1.143$ respectively. The results of Gaussian profile fitting to each feature are given in Table \ref{spec-red-tab}. There are also two other possible perplexing features in the spectrum; an emission line at 5618~{\AA} and an absorption line at 4785~{\AA}. The source of these features is unknown as neither are associated with the lensing galaxy or the 5995~{\AA} emission line from the background source. Unfortunately, only a single exposure of CLASS B0850+054 was taken through LRIS-B (see Section \ref{spec-obs}). Therefore, the signal-to-noise ratio in this region of the spectrum is very low and no concrete conclusions regarding the detection or source of these two confusing features can be made. Further observations are required. Nevertheless, a secure lensing galaxy redshift of $z_{l}=0.5883\pm0.0006$ has been obtained.

\begin{table*}
\begin{center}
\begin{tabular}{cllccl} \hline
Source		& \multicolumn{1}{c}{Lens/}  & \multicolumn{1}{c}{Line} & \multicolumn{1}{c}{$\lambda_{\mbox{0}}$}  & \multicolumn{1}{c}{$\lambda_{\mbox{obs}}$} & \multicolumn{1}{c}{$\langle z \rangle$} \\
		& \multicolumn{1}{c}{Source} & \multicolumn{1}{c}{ID}   & \multicolumn{1}{c}{{[\AA}]}     & \multicolumn{1}{c}{[\AA]} &   \\ \hline
CLASS B0128+437	& Lens	 & [O~{\sc ii}]?	& 3727 & 7994.8    & 1.145?\\  
		&	 & H${\beta}$?    	& 4861 & 	   & 0.645?\\		
		&        & H$\alpha$?   	& 6563 & 	   & 0.218?\\
		& Source & Ly$\alpha$   	& 1216 & 5014.8    & $3.1240\pm0.0042$\\
		&	 & C~{\sc iv}?   	& 1549 & 6383.5    & \\	
CLASS B0445+123	& Lens   & H$\eta$ 		& 3835 & 5974.7    & $0.5583\pm0.0003$\\
		&        & H$\zeta$ 		& 3889 & 6056.6    & \\
		&        & Ca~{\sc ii} K 	& 3934 & 6126.9    & \\
                &	 & Ca~{\sc ii} H 	& 3968 & 6183.2    & \\
		&	 & G-Band		& 4304 & 6712.7    & \\
		&	 & H$\gamma$		& 4340 & 6764.1    & \\
		&	 & Mg b			& 5174 & 8066.4    & \\
		& Source & No features  	&      &           & \\
CLASS B0631+519 & Lens-1 & [O~{\sc ii}]  	& 3727 & 4061.3    & $0.0896\pm0.0001$\\
		&	 & [Ne~{\sc iii}]	& 3869 & 4216.0    & \\
		&	 & H$\delta$		& 4101 & 4467.8    & \\
 		&	 & H$\gamma$		& 4340 & 4730.5    & \\
		&	 & H$\beta$		& 4861 & 5297.0    & \\
		&	 & [O~{\sc iii}]	& 4959 & 5402.7    & \\
		&	 & [O~{\sc iii}]	& 5007 & 5455.5    & \\
		&	 & H$\alpha$		& 6563 & 7149.6    & \\
		& Lens-2 & BL3096       	& 3096 & 5012.4    & $0.6196\pm0.0004$\\
		&        & BL3580       	& 3580 & 5802.9    & \\
		&        & BL3835		& 3835 & 6203.8    & \\
		&        & Ca~{\sc ii} K 	& 3934 & 6372.0    & \\
		&        & Ca~{\sc ii} H 	& 3968 & 6429.2    & \\
		&        & G-Band		& 4304 & 6973.5    & \\
		&        & H$\gamma$    	& 4340 & 7029.6    & \\
		& Source & Ly$\alpha$?  	& 1216 & 5039.7    & 3.144?\\
		&        & Mg~{\sc ii}?  	& 2798 &	   & 0.801?\\
CLASS B0850+054 & Lens   & BL3835	      	& 3835 & 6088.6    & $0.5883\pm0.0006$\\
		& 	 & Ca~{\sc ii} K 	& 3934 & 6242.3    & \\
		& 	 & Ca~{\sc ii} H 	& 3968 & 6295.3    & \\
		& 	 & Ca4455		& 4455 & 7083.9    & \\
		& 	 & Fe4531		& 4531 & 7201.0    & \\
		& 	 & Mg b			& 5174 & 8222.6    & \\
		& Source & Ly$\alpha$?  	& 1216 & 5995.4    & 3.930? \\ 
		&        & Mg~{\sc ii}?  	& 2798 &           & 1.143? \\ \hline
\end{tabular}
\caption{The results of Gaussian line profiles fitted to the un-smoothed optical spectra of CLASS gravitational lenses.}
\label{spec-red-tab}
\end{center}
\end{table*}

\section{Discussion}
\label{spec-disc}

\subsection{Properties of the Lensing Galaxies}

It is possible to estimate the mass within the Einstein Radius of each gravitational lens system using the newly obtained lens and source redshifts. Furthermore, our knowledge of the global properties of early/late-type galaxy populations can be used to place some constraints on the unknown redshifts in each of the gravitational lenses studied here. The mass contained within the Einstein Radius, $\theta_{E}$ ($\sim \Delta \theta/2$ where $\Delta \theta$ is the image separation), is defined as,
\begin{equation}
M_{E} \approx 1.24 \times 10^{11} \left(\frac{\theta_{E}}{1\arcsec}\right)^2 \left(\frac{D}{\mbox{1 Gpc}}\right) M_{\odot},
\end{equation}
where 
\begin{equation}
D=\frac{D_{l}D_{s}}{D_{ls}},
\end{equation}
 and $D_{l}$ is the angular-diameter distance to the lens, $D_{s}$ is the angular-diameter distance to the source and $D_{ls}$ is the angular-diameter distance from the lens to the source. Studies of other CLASS gravitational lenses have found the mass enclosed within the Einstein Radius to be typically $\sim10^{10}-10^{11}~h^{-1}~M_{\odot}$ for early-type lensing galaxies and $\sim 10^{10}~h^{-1}~M_{\odot}$ for late-type lensing galaxies. However, it should be noted that in some cases the lensing potential will be supplemented by the presence of a group or cluster of galaxies supporting the main deflector (\citealt*{keeton00}; \citealt{tonry00,fassnacht02}).

For each of the gravitational lenses studied here either the lensing galaxy or the lensed source redshift is unknown. For CLASS B0128+437 and CLASS B0850+054 a single emission line was detected from the lensing galaxy and lensed source, respectively. Therefore, the possible identifications of these single lines, given in Table \ref{spec-red-tab}, will be used in the following analysis. For both CLASS B0445+123 and CLASS B0631+519 there were no firm detections of emission lines from the background source. Therefore, a conservative redshift of $z_{s}=2.0\pm1.0$ is adopted for the lensed source. The physical size of the Einstein Radius ($r_{E}$) and resulting mass enclosed within the Einstein Radius for each gravitational lens are given in Table \ref{lens-z-data}. 

The single emission line from the lensing galaxy in CLASS B0128+437 is unlikely to be H$\alpha$ as it results in a low mass of $M_{E} = 0.53 \times 10^{10}~h^{-1}~M_{\odot}$ within the Einstein Radius. Alternatively, both the H$\beta$ and [O~{\sc ii}] identifications are consistent with a late-type lensing galaxy, with an enclosed mass of $M_{E} = 1.41\times 10^{10}~h^{-1}~M_{\odot}$ and $M_{E} = 2.46\times 10^{10}~h^{-1}~M_{\odot}$ respectively. The H$\beta$ redshift is in the region of the canonical lensing galaxy redshift of $z_{l} \sim 0.6$. However, as CLASS B0128+437 is very faint in the optical ($I\sim 25$) the higher [O~{\sc ii}] redshift should not be discounted. Furthermore, [O~{\sc ii}] is a commonly observed emission line in late-type galaxies \citep{kennicutt92}. In the cases of CLASS B0445+123 and CLASS B0631+519 a source redshift of $z_{s} = 1 - 3$ required an enclosed mass within the Einstein Radius of $M_{E} = 6-14\times 10^{10}~h^{-1}~M_{\odot}$, which is consistent with an early-type lensing galaxy. Therefore, no limit can be placed on the source redshift from this analysis. The Ly$\alpha$ and Mg~{\sc ii} candidate identifications of the emission line from the background source in CLASS B0850+054 resulted in an enclosed mass of $M_{E} = 1.97\times 10^{10}~h^{-1}~M_{\odot}$ and $M_{E} = 3.40\times 10^{10}~h^{-1}~M_{\odot}$ respectively. As the CLASS B0850+054 deflector is a late-type galaxy, neither line identification can be ruled out at this stage.

\begin{table*}
\begin{center}
\begin{tabular}{cllrllcr} \hline
Gravitational	& \multicolumn{1}{c}{$z_{l}$}	        & \multicolumn{1}{c}{$z_{s}$}	& \multicolumn{1}{c}{$D_{l}$}	& \multicolumn{1}{c}{$D_{s}$}	& \multicolumn{1}{c}{$D_{ls}$}	& $r_{E}$	& \multicolumn{1}{c}{$M_{E}$}	\\
Lens		&		    &		        & \multicolumn{1}{c}{[$h^{-1}$Mpc]}		& \multicolumn{1}{c}{[$h^{-1}$Mpc]}		& \multicolumn{1}{c}{[$h^{-1}$Mpc]}		& [$h^{-1}$kpc]	  & \multicolumn{1}{c}{[$10^{10}h^{-1}M_{\odot}$]}\\ \hline
CLASS B0128+437	& $0.218\pm0.002$   & $3.1240\pm0.0042$ & $509.3\pm3.6$		& $1098.6\pm0.5$ & $948.1\pm1.6$ & 0.7		& $0.53\pm0.01$	\\
		& $0.645\pm0.002$   & 		        & $997.1\pm1.3$		&		 & $700.8\pm1.1$ & 1.3		& $1.41\pm0.03$	\\
		& $1.145\pm0.002$   &		        & $1188.6\pm0.4$	&		 & $480.3\pm0.9$ & 1.6		& $2.46\pm0.05$	\\ \hline
CLASS B0445+123 & $0.5583\pm0.0003$ & $1.0$ 	        & $932.6\pm0.2$		& 1156.3	 & 429.7	 & 3.0		& $13.56\pm0.27$\\
		& 		    & $2.0$	        &			& 1208.6	 & 723.2	 &		& $8.42\pm0.17$	\\
		& 		    & $3.0$	        & 			& 1112.2	 & 748.9	 &		& $7.48\pm0.15$	\\ \hline
CLASS B0631+519 & $0.6196\pm0.0004$ & $1.0$ 	        & $979.6\pm0.3$		& 1156.3	 & 363.1	 & 2.8		& $13.01\pm0.26$\\
		&		    & $2.0$	        &			& 1208.6	 & 679.8	 &		& $7.26\pm0.15$	\\
		&		    & $3.0$	        &			& 1112.2	 & 715.6	 &		& $6.36\pm0.13$	\\ \hline
CLASS B0850+054 & $0.5883\pm0.0006$ & $1.143\pm0.002$   & $956.4\pm0.5$		& $1188.2\pm0.4$ & $479.4\pm1.6$ & 1.6		& $3.40\pm0.07$	\\
		& 		    & $3.930\pm0.002$   & 			& $1011.1\pm0.2$ & $703.0\pm0.3$ &		& $1.97\pm0.04$	\\
\hline
\end{tabular}
\caption{The properties of the lensing galaxies in CLASS B0128+437 ($\Delta\theta=0.54\arcsec$), CLASS B0445+123 ($\Delta\theta=1.32\arcsec$), CLASS B0631+519 ($\Delta\theta=1.16\arcsec$) and CLASS B0850+054 ($\Delta\theta=0.68\arcsec$) based on the observations presented in this paper. For CLASS B0128+437 and CLASS B0850+054 the respective lens and source redshifts are based on the single emission line detected in each spectrum. For CLASS B0445+123 and CLASS B0631+519 a typical background source redshift of $z_{s}=1-3$ has been used. An $\Omega_{M}=0.3$, $\Omega_{\Lambda}=0.7$ flat-universe, with a Hubble parameter $H_{0}=100~h$~km~s$^{-1}$~Mpc$^{-1}$ has been assumed during this analysis.}
\label{lens-z-data}
\end{center}
\end{table*}

\subsection{The Low-$\bmath z$ Lensing Galaxy in CLASS B0631+519}

An intriguing feature of the CLASS B0631+519 spectrum, shown in Fig. \ref{lens-0631}, is the presence of a low redshift galaxy at $z = 0.0896$. In the analysis presented above, this galaxy has been assumed to be one of two lensing galaxies. Here, we investigate the single lensing galaxy scenario by determining whether a $z=0.0896$ deflector is capable of producing $1.16\arcsec$ image-splitting of a $z=0.6196$ background source.

Single deflectors at low redshifts are uncommon, but nevertheless do exist. For example, MG1549+3047 ($z_{l}=0.11$; \citealt{Lehar93}) and Q2237+030 ($z_{l}=0.04$; \citealt{huchra85}) have low redshift lensing galaxies. However, HST imaging of MG1549+3047 \citep{kochanek00} and Q2237+030 \citep{falco99} has also shown both to have highly luminous ($V \sim 15-16$-mag) and therefore massive lensing galaxies. Unfortunately, the resolution of the York et al. WHT $R$-band image was insufficient to resolve the individual lensing galaxies. Therefore, only an upper-limit can be placed on the luminosity of the nearby CLASS B0631+519 lensing galaxy by i) assuming it is responsible for all of the $21.4$-mag $R$-band emission, ii) converting this emission to the rest-frame $B$-band magnitude, $m_{B}$, using the galaxy colour differences and non-evolutionary $k$-correction, $k(z)$, of \citet*{coleman80}, and iii) comparing with the luminosity of an $L^*$ galaxy \citep{norberg02}. The absolute $B$-band magnitude, $M_{B}$, was calculated using,
\begin{equation}
M_{B} = m_{B} - 5 \log_{10} \left( \frac{D_{L}}{10~\mbox{pc}} \right) - k(z),
\end{equation}
where $D_{L}$ is the luminosity distance. This was found to be $M_{B} > -14.4\pm0.1 + 5\log_{10}h$ which corresponds to a luminosity of $L_{B} < 9.04 \pm 0.8 \times 10^{7}~h^{-2}~L_{\odot}$ or $L_{B} < L^{*}/94$. Assuming that this galaxy is producing the $\Delta \theta = 1.16\arcsec$ image splitting of the distant $z=0.6196$ galaxy results in a mass enclosed within the Einstein Radius of $M_{E} = 1.21\pm0.02 \times 10^{10}~h^{-1}~M_{\odot}$ and a mass-to-light ratio of $(M/L)_{B} \ga 134~h~(M/L)_{\odot}$. However, it is clear from the spectrum of CLASS B0631+519 that the $z=0.6196$ galaxy is the dominant source of emission in $R$-band. Therefore, this calculation provides a firm lower limit of the mass-to-light ratio of the low redshift galaxy. As the typical mass-to-light ratios of galaxy-scale deflectors are at most only a few tens of $(M/L)_{\odot}$ and there is currently no evidence that the lensing is cluster assisted, this arrangement appears to be unlikely. However, the possibility that the low redshift galaxy is a dark or dust enshrouded lens cannot be ruled out at this stage. There is currently only one `dark lens' candidate, CLASS B0827+525\footnote{The binary quasar hypothesis has not been ruled out for this system.} \citep{koopmans00a}, in the entire CLASS gravitational lens sample, which along with CLASS B0631+519 would result in a CLASS `dark lens' fraction of $\la 10$~per~cent. Alternatively, the narrow emission line spectrum of the low redshift object is consistent with a nuclear starburst galaxy which has undergone or is currently undergoing a period of intense star-formation. Therefore, it is possible that the under-luminous nature of the low redshift galaxy could be due to extinction caused by the presence of dust. Further observations would need to be conducted to investigate this possibility. Whether CLASS B0631+519 is a gravitational lens with two lensing galaxies or a single dark/dusty lensing galaxy is unclear. However, it is most likely that the $z = 0.6196$ galaxy is the main deflector and the $z = 0.0896$ object is a dwarf star-forming galaxy which happens to be coincident along the line-of-sight.

\section{Conclusions}
\label{spec-conc}

We have presented new optical spectra of four recently discovered gravitational lenses from CLASS. The redshift of the background source in CLASS B0128+437 was found to be $z_s=3.1240\pm0.0042$ and the lensing galaxy redshifts in CLASS B0445+123 and CLASS B0850+054 were found to be $z_{l}=0.5583\pm0.0003$ and $z_{l}=0.5883\pm0.0006$ respectively. Interestingly, we discovered that CLASS B0631+519 may have two lensing galaxies. One is a low luminosity galaxy at $z_{l,1}=0.0896\pm0.0001$ which has little effect on the lensing potential, and the other is the primary lensing galaxy at $z_{l,2}=0.6196\pm0.0004$. The other possibility is that the low redshift galaxy is a dark/dust enshrouded gravitational lens. However, the scenario in which the main lens is at $z = 0.6196$ is currently favoured. We find the lensing galaxies in CLASS B0445+123 and CLASS B0631+519 ($l,2$) to be early-type galaxies, with Einstein Radii of $2.8$ and $3~h^{-1}$~kpc respectively. The lensing galaxy in CLASS B0850+054 was found to be consistent with a late-type galaxy with an Einstein Radius of 1.6~$h^{-1}$~kpc.

For the redshift data presented in this paper to be useful for studying galaxy formation at high redshift or investigating the cosmological parameters, they must be coupled with the redshifts of the missing components. We have identified isolated emission lines in the CLASS B0128+437 and CLASS B0850+054 spectra which we hypothesised identifications for. However, these tentative identifications will only be confirmed by further observations. Of the 22 gravitational lenses discovered by CLASS, 17 have measured lensing galaxy redshifts and 12 have background source redshifts recorded. Obtaining spectra of the remaining gravitational lenses with unknown source or lens redshifts is an essential step in the follow-up of the CLASS gravitational lensing survey. To achieve this goal, further deep spectroscopic observations are required in the optical and near-infrared. 

\section*{Acknowledgments}
The data presented herein were obtained at the W. M. Keck Observatory, which is operated as a scientific partnership among the California Institute of Technology, the University of California and the National Aeronautics and Space Administration. The Observatory was made possible by the generous financial support of the W. M. Keck Foundation. The authors wish to recognise and acknowledge the very significant cultural role and reverence that the summit of Mauna Kea has always had within the indigenous Hawaiian community.  We are most fortunate to have the opportunity to conduct observations from this mountain. This research has made use of the NASA/IPAC Extragalactic Database (NED) which is operated by the Jet Propulsion Laboratory, California Institute of Technology, under contract with the National Aeronautics and Space Administration. JPM acknowledges the receipt of a PPARC studentship. LVEK acknowledges the support from an STScI Fellowship grant. RDB is supported by an NSF grant AST--0206286.

\label{lastpage}

\end{document}